\documentclass[prl, twocolumn, amsmath, nofootinbib, floatfix, showpacs,
  preprintnumbers,letter]{revtex4}  

\usepackage{graphicx}
\usepackage{xcomment}
\usepackage{color,aas_macros}

\newcommand{\rf}[1]{\ref{fig:#1}}

\def\lsim{\mathrel{\raise.3ex\hbox{$<$\kern-.75em\lower1ex\hbox{$\sim$}}}}
\def\gsim{\mathrel{\raise.3ex\hbox{$>$\kern-.75em\lower1ex\hbox{$\sim$}}}}

\def\cmm2{{\,\rm cm^{-2}}}
\def\cm2{{\,{\rm cm}^2}}
\def\cmm3{{\,{\rm cm}^{-3}}}
\def\gcmm3{{\,{\rm g\,cm^{-3}}}}

\def\fun#1#2{\lower3.6pt\vbox{\baselineskip0pt\lineskip.9pt
  \ialign{$\mathsurround=0pt#1\hfil##\hfil$\crcr#2\crcr\sim\crcr}}}

\def\be{\begin{equation}}
\def\ee{\end{equation}}
\def\bea{\begin{eqnarray}}
\def\eea{\end{eqnarray}}
\newcommand{\vs}{\nonumber\\}

\newcommand{\ec}[1]{Eq.~(\ref{eq:#1})}

\newcommand{\eql}[1]{\label{eq:#1}}

\def\be{\begin{equation}}
\def\ee{\end{equation}}
\def\bea{\begin{eqnarray}}
\def\eea{\end{eqnarray}}

\newcommand\kappam{\kappa^{\rm obs}}
\newcommand\kappat{\kappa^{\rm true}}
\newcommand\bias{b_m}
\newcommand\mobs{m^{\rm obs}}
\newcommand\sobs{s^{\rm obs}}
\newcommand\psh{\mathcal{P}}

\begin{document}

\preprint{FERMILAB-PUB-10-396-T}

\title{Magnification as a Tool in Weak Lensing}


\author{Alberto Vallinotto$^1$, Scott Dodelson$^{1,2,3}$, Pengjie Zhang$^4$}
\affiliation{$^1$Center for Particle Astrophysics, Fermi National
Accelerator Laboratory, Batavia, IL~~60510}
\affiliation{$^2$Department of Astronomy \& Astrophysics, The
University of Chicago, Chicago, IL~~60637}
\affiliation{$^3$Kavli Institute for Cosmological Physics, Chicago, IL~~60637}
\affiliation{$^4$Key Laboratory for Research in Galaxies and Cosmology, Shanghai Astronomical Observatory, Nandan Road 80, Shanghai, 200030, China}
\date{\today}

\begin{abstract}
Weak lensing surveys exploit measurements of galaxy ellipticities. These
measurements are subject to errors which degrade the cosmological information
that can be extracted from the surveys. Here we propose a way of using the galaxy
data themselves to calibrate the measurement errors. In particular, the
cosmic shear field, which causes the galaxies to appear elliptical, also changes
their sizes and fluxes. Information about the sizes and fluxes of the galaxies
can be added to the shape information to obtain more robust information about
the cosmic shear field. The net result will be tighter constraints on
cosmological parameters such as those which describe dark energy.  
\end{abstract}

\maketitle

\textit{Introduction}. Weak gravitational lensing~\cite{Bartelmann:1999yn, Schneider:2005ka, Munshi:2006fn} has the potential to probe some of the most outstanding problems in cosmology. Hidden in the pattern of
the ellipticities of background galaxies is information about the cosmic shear
field, which in turn depends on the large scale properties of the universe,
including the nature of the dark energy~\cite{Caldwell:2009ix,Frieman:2008sn}. 
Among the systematic hurdles that
must be overcome in order to mine this information is bias in the measurements
of these ellipticities. Here we focus on {\it multiplicative bias} 
\cite{Huterer:2005ez,Heymans:2005rv,Massey:2006ha,Amara:2007as}, the fact that
the observed ellipticity of a galaxy is related to its true ellipticity via a
multiplicative factor that is not necessarily equal to one. When ellipticities are
converted to convergences (where $\kappa$ is the convergence along the line of
sight to the background galaxies), this translates to 
\begin{equation}
\kappam(\hat n,z) = \bias(z) \kappat(\hat n,z)
\eql{bias}
\end{equation}
where the assumption that the multiplicative bias $\bias$ depends only on the redshift of the
background probes follows the arguments of
Refs.~\cite{Huterer:2005ez,Heymans:2005rv,Massey:2006ha,Amara:2007as} that the
angular dependence leads only to smaller, higher-order corrections. 

Here we propose a method of calibrating the multiplicative bias, exploiting
the effect that weak lensing has on the sizes and fluxes of the
background galaxies. While shapes are usually used to infer $\kappa$, sizes
and fluxes are also distorted. As such, these observables carry information about the
convergence field\footnote{The distortions in size and magnitude also introduce selection bias 
\cite{Schmidt:2009rh,Schmidt:2009ri}, which will bias cosmological results if not accounted for properly.} and can also be used to improve cluster mass estimates \cite{Rozo_Schmidt:unpub}. We show here that future surveys may be able to use
this information to calibrate the multiplicative bias, thereby enabling us to capture more of the information contained in the cosmic shear field. For concreteness we focus mainly on two upcoming 
lensing surveys: the Dark Energy
Survey~\cite{Abbott:2005bi} (DES) and the Large Synoptic Survey Telescope~\cite{Ivezic:2008fe} (LSST), both at a single redshift slice (so that $\bias$
is constant).

\textit{The Impact of Lensing on Sizes and Fluxes}.
Weak lensing increases the size of a given galaxy by a factor of $1+\kappa$
and the flux by $1+2\kappa$, 
corresponding to a decrease in magnitude by $2.5\ln(1+2\kappa)/\ln(10)$. The
average size and flux of galaxies in 
a survey, however, are affected by lensing in a more complex way due to the thresholds
for inclusion in the survey. Consider Fig.~\rf{distributions+lensing}, which 
depicts the size and $i$-magnitude distributions of galaxies observed in the 
Hubble-GOODS survey~\cite{:2003ig} as they might appear in DES behind a region
with $\kappa=0.1$. Cuts in size and magnitude
are depicted by the vertical lines and a mean seeing of $0.9''$ has been added to the sizes.
Although each individual galaxy increases in size/magnitude, the mean size/magnitude is also affected
by the small/faint galaxies that are promoted into the survey by lensing. Thus, the change
in the mean size/magnitude is not given by the simple relationships above.

\begin{figure*}[ht]
\includegraphics[width=0.9\columnwidth]{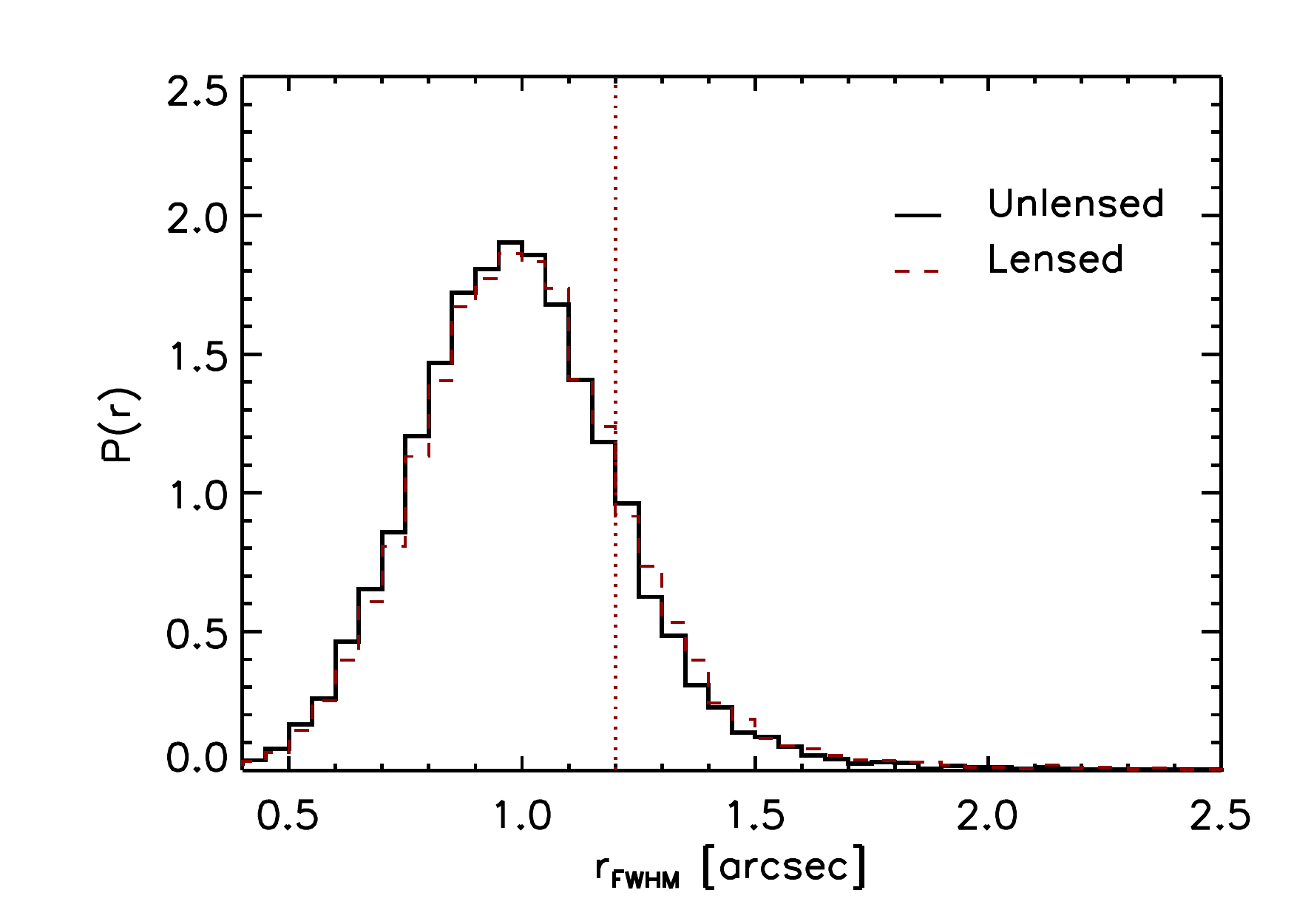}
\includegraphics[width=0.9\columnwidth]{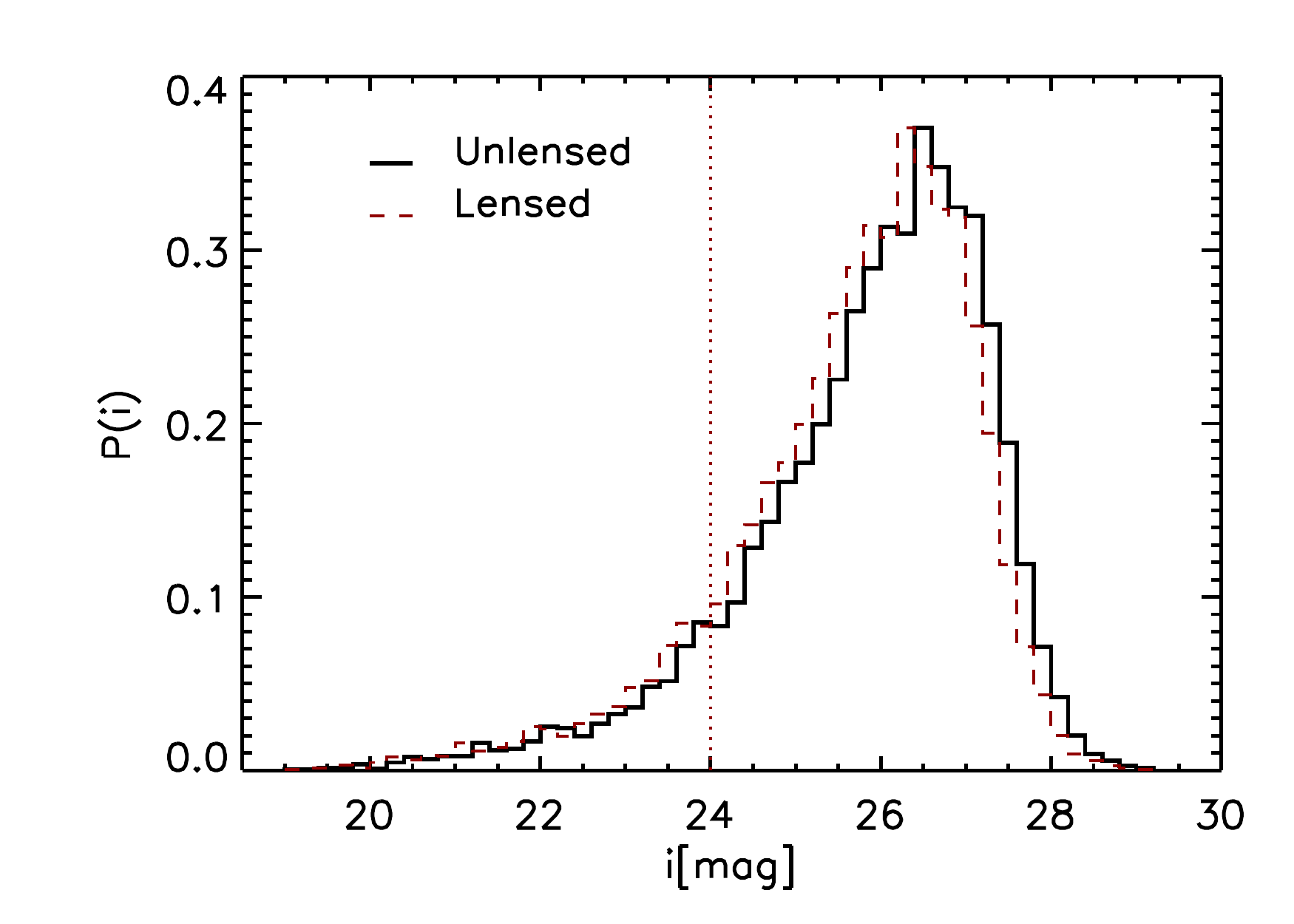}
\caption{\label{fig:distributions+lensing} Distributions of the sizes (left
  panel) and the magnitudes (right panel) for a DES-like survey with (dashed)
  and without (solid) lensing by a convergence $\kappa=0.1$. Regions with
  positive $\kappa$ will therefore have larger and brighter galaxies. Vertical
  lines show the fiducial thresholds for the DES survey.} 
\end{figure*} 

We quantify the effect of lensing of the mean sizes and magnitudes by introducing coefficients
$g_s,g_m$ defined via
\begin{eqnarray}
\langle\mobs_i\rangle &\equiv& m_0+g_m\kappat_i  \qquad {\rm [magnitude]}\vs
\langle\sobs_i\rangle &\equiv& s_0 (1+g_s \kappat_i) \qquad {\rm [size]}.
\eql{qs}
\end{eqnarray}
Here $m_0$ and $s_0$ are the mean magnitude and size in the entire galaxy sample, and the index 
$i$ labels an angular pixel in which there are many galaxies, all of which are affected
by the same convergence $\kappat_i$. Since $s_0$ and $m_0$ are obtained by averaging over all galaxies
in the survey (behind regions with both positive and negative $\kappa$), they can
be determined very accurately. The mean size and magnitude of
background galaxies in a pixel, therefore, contain information about the
cosmic shear field affecting that pixel.

To extract this information from sizes and magnitudes, we
need to know the coefficients $g_m$ and $g_s$. One hint comes from 
the recent detection of lensing on
the average flux of SDSS quasars~\cite{Menard:2009yb}. They found a value of $g_m=-0.25$ ($-C_S$ in their
notation). 
Using the deep HST data, we can simulate DES/LSST conditions and
estimate the coefficients. We take the unlensed
sizes and magnitudes, adopt a value of $\kappa$, generate a new set of simulated data by
$m\rightarrow m-2.5\ln (1+2\kappa)/\ln 10$ and $s\rightarrow s(1+\kappa)$, and then apply simulated
seeing and cuts
of $\sobs>1.2''$ and $\mobs < 24$. The resulting means $\langle m^{\rm obs}\rangle$ and
$\langle s^{\rm obs}\rangle$ then determine the $g$'s via, e.g., $g_m = (\langle \mobs\rangle -m_0)/\kappa$.
We find $g_m=-0.3$ and $g_s=0.25$, fairly independently of cuts and $\kappa$, so we use these values in the projections.

Estimating $g_m$ and $g_s$ from survey data is likely to be more difficult.  
The standard estimate for $g_m$, e.g., is 
\begin{equation}
g_m=\left[1-\frac{m_cN(m_c)}{\int_0^{m_c}mN(m)dm}+\frac{N(m_c)}{\int_0^{m_c}N(m)dm}\right]\frac{5}{\ln
  10} ,\end{equation}
where $N(m)dm$ is the number of galaxies in the magnitude interval $m$-$m+dm$ and $m_c$
is the magnitude cut.
This formula though neglects the real world complexities introduced by multiple cuts (size and magnitude) and
the finite statistics in the magnitude bin used to estimate $N(m)$. Instead, we may need to use data from even deeper surveys to calibrate
the $g$'s for the survey of interest, just as have used used HST data here to estimate the $g$'s for DES.


%
%

%

%


\textit{Reducing Multiplicative Bias.} We now envision using all three sets of
observables (ellipiticites, sizes, and magnitudes), each of which depends on $\kappat$, to constrain the multiplicative bias, and therefore reduce the errors on cosmological parameters. As an illustration, we consider the 
case where there is only a single cosmological parameter, the amplitude of the power spectrum of the convergence, $P_\kappa$.
If we had no information about the bias $\bias$, then ellipticity measurements
alone could not determine  the amplitude of the power spectrum,
$P_\kappa$. Technically, if the amplitude of the power spectrum were
characterized by $a$, with $a=1$ being the true value, there would be a
complete degeneracy between $a$ and $\bias$, since 
\begin{eqnarray}
\langle \kappam_i \kappam_j \rangle &=&  \bias^2 \langle \kappat_{i} \kappat_{j} \rangle
+\delta_{ij} \sigma_\kappa^2 \vs  
&=& a^2 \bias^2 \int \frac{d^2l}{(2\pi)^2} \psh_\kappa(l) J_0(l\theta_{ij})
+\delta_{ij} \sigma_\kappa^2\vs 
&\equiv & a^2\bias^2\xi_\kappa(\theta_{ij})+\delta_{ij} \sigma_\kappa^2 
\eql{kk}
\end{eqnarray}
where $\theta_{ij}$ is the angular distance between the two pixels, $\psh_\kappa$ is
the (assumed known) shape of the power spectrum ($P_\kappa=a\psh_\kappa$), and
$\sigma_\kappa$ is the rms of the ellipticities in the absence of a signal, due to
shape noise and measurement errors. Observations of ellipticities then depend
only on the product $a\bias$, so there is a complete degeneracy between these two parameters.
  
The sizes and magnitudes contain information that break this degeneracy. A
simple way to exploit this information is to consider the full set of
two-point functions of (convergence, size, and magnitude) for each pair of
pixels:  
\begin{eqnarray}
\langle \kappam_{i} (\sobs_{j}-s_0) \rangle&=& a^2 \bias g_s s_0 \xi_\kappa(\theta_{ij}),\\
\langle \kappam_{i} (\mobs_{j}-m_0) \rangle&=&  a^2 \bias g_m \xi_\kappa(\theta_{ij}),\\
\langle (\sobs_{i}-s_0) (\sobs_{j}-s_0) \rangle&=&
\delta_{ij}\sigma_s^2\left[1+a^2g_s^2\xi_\kappa(\theta=0)\right] \vs 
&+&a^2 g_s^2 s_0^2 \xi_\kappa(\theta_{ij}),\eql{ss}\\
\langle (\sobs_{i}-s_0) (\mobs_{j}-m_0) \rangle &=&a^2g_m g_s s_0 \xi_\kappa(\theta_{ij}),\\
\langle (\mobs_{i}-m_0) (\mobs_{j}-m_0) \rangle&=& \delta_{ij}\sigma_m^2
+a^2 g_m^2\xi_\kappa(\theta_{ij}).\eql{mm}
\end{eqnarray}
The variances $\sigma_s^2$ and $\sigma_m^2$ include contributions from
intrinsic scatter in sizes and magnitudes and also the measurement errors
expected in the survey. Here, we have assumed that the intrinsic sizes and fluxes of
galaxies are uncorrelated with one another. 

The data set will then contain $3N$ numbers: the average size, magnitude, and shear/convergence of all galaxies in a set of $N$ pixels. 
To assess how powerful this information will be, we construct the Fisher matrix. 
The $2\times2$ Fisher matrix is 
\begin{equation}
F_{\alpha\beta}=\frac{1}{2}{\rm Tr}\left[C_{,\alpha}C^{-1}C_{,\beta}C^{-1}\right].\label{Fisher}
\end{equation}
where $\alpha,\beta$ run over the two parameters $a$ and $\bias$, the trace is
over all the $3N$ observables and $C$ is their $3N\times 3N$ covariance matrix with elements given in Eqs.~(\ref{eq:kk}-\ref{eq:mm}). 


DES (LSST) will cover about $5,000\, (20,000)$ sq.~degrees, and we consider pixels of size
$\Delta\theta^2=10$ square arcmin. The total number of pixels is then
$=1.8\times10^6\, (7.2\times 10^6)$. We focus only on the redshift range
$z\in[0.9,1.1]$ with an expected number of galaxies per pixel~of $M=15 \,(100)$.  
The parameters assumed for magnitudes and sizes are summarized in
Tab.~\ref{table0}. 

\begin{table}[t]
\begin{center}
\begin{tabular}{cccccc}
\hline
Probe & Mean  & Dispersion & $g$ & Cut \\
\hline
\hline
Size (arcsec)& 1.5 (0.9) & 0.33  (0.266) & 0.25 & 1.2 (0.7) \\
Magnitude & 22.7 (24.4) & 1.1 (1.26)   & -0.3 & 24 (26)\\
\hline
\end{tabular}
\end{center}
\caption{Assumed values for sizes and magnitudes of DES and LSST. Values for
  the sizes and magnitudes of \textit{single} galaxies for the DES survey, so
  the scatter in a single pixel containing $M=15$ galaxies is
  $\sigma_s=0.33''/\sqrt{15}$ and $\sigma_m=1.1/\sqrt{15}$. Values in parentheses are those adopted
for LSST, with $M=100$ assumed.} 
\label{table0}
\end{table}%

Fig.~\rf{Ellipse} shows the projected constraints from DES assuming that all
the pixels are uncorrelated (so the Fisher matrix is simply the one-pixel
Fisher matrix times $1.8\times10^6$). The figure shows that multiplicative
bias in a single redshift bin can be pinned down at the 5\% level with the aid of size and magnitude
measurements. Also shown is the projection for LSST. Here the requirements on the bias will be more severe because the
statistical power is much higher~\cite{Amara:2007as}, and indeed the extra
information does pin down multiplicative bias at the percent level.

The projected errors in Fig.~\rf{Ellipse} retain the degeneracy between $a$ and $\bias$ 
that afflicts the shear-only measurements. This is an indication that the shear measurements carry the most
statistical weight. To confirm this, consider the signal to noise of the shear measurement. Taking the ratio of the
two terms on the right in \ec{kk} and weighting by the number of galaxies $M$ in a pixel leads to $(S/N)_{\rm shear} \simeq\sqrt{M\xi/\sigma_\kappa^2}$. For 
DES (LSST) this is of order 0.1 (0.3). By contrast, a similar estimate for the size measurement using \ec{ss} leads to $(S/N)_{\rm size} \simeq
\sqrt{Mg_s^2s_0^2\xi/\sigma_s^2}$, or 0.04 (0.07) for DES (LSST). Although the dispersion in sizes is comparable to shape noise, the signal is suppressed by a factor of $g_s$. A similar estimate for the magnitudes yields even smaller signals. So the shear measurements dominate the constraints, and the utility of the size/magnitude measurements is to break the degeneracy between the amplitude of the clustering ($a$) and the multiplicative bias ($\bias$).

We have neglected correlations between pixels. When
these are added in, the constraints will become
tighter. We have not computed the Fisher matrix including all 
correlations, but we have studied how the constraints with and without
correlations compare for smaller fields as the number of pixels increases
. These studies suggest that there is additional information
in the correlations which will further tighten the constraints on
multiplicative bias by at least 10\%.  

\begin{figure}
\includegraphics[width=0.4\textwidth]{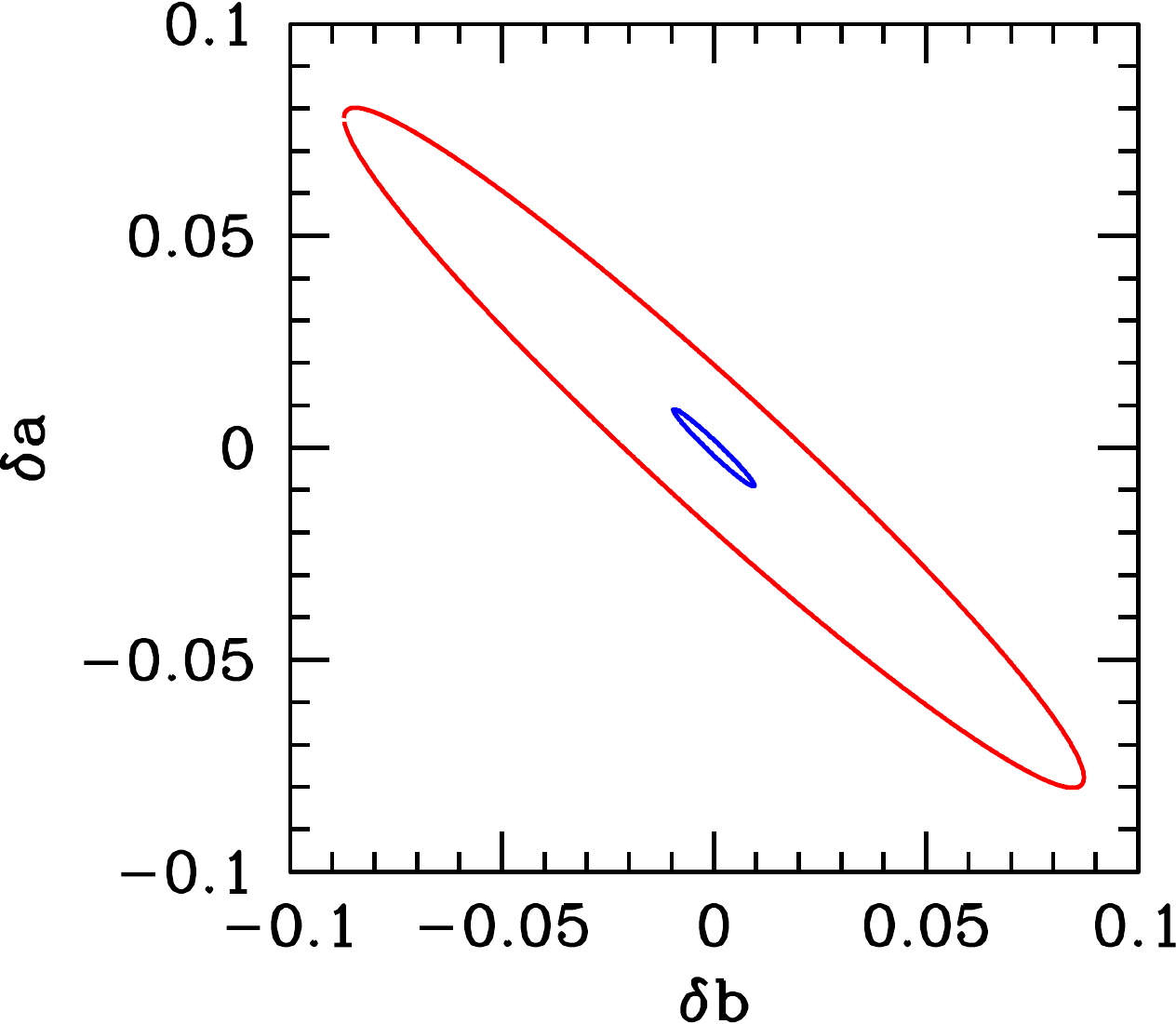}
\caption{\label{fig:Ellipse} Projected constraints on the multiplicative bias
$\bias$ and amplitude of the power spectrum $a$ for two surveys. Outer contour
 shows the projection for DES (5000 square degrees) with 1.5 galaxies
per square arcminute in the redshift range $[0.9,\,1.1]$. Inner contour shows
projections for an LSST-like survey with 10 galaxies per square arcminute in this redshift range and
 20,000 square degrees.} 
\end{figure}

\textit{Conclusions}. Weak lensing affects several observed properties of galaxies: 
not only does it distort their shapes, but it also alters their observed sizes and
magnitudes.  We have demonstrated that these other distortions can turned into an asset:
the lensing effect on the average
galaxy size and magnitude helps to constrain multiplicative bias. 
The comprehensive way to determine how successful these new observables will
be at improving cosmological constraints is to add them to the program initiated by
Bernstein~\cite{Bernstein:2008aq}, where the Fisher matrix for all parameters
(cosmological and nuisance) is determined for a fixed set of
measurements of $\kappa$ and the density of sources. Here we have estimated the improvement
in a simple setting where only the amplitude of the power spectrum is unknown. This simple
example suggests that the added information is potentially useful and should be incorporated into
the more comprehensive program and ultimately into the full analysis pipeline of upcoming surveys.

\textit{Acknowledgments}.
We thank Eduardo Rozo and Fabian Schmidt for useful comments and discussions. 
This work has been supported by the US Department of Energy, including grant
DE-FG02-95ER40896, and by National Science Foundation Grant AST-0908072. PJZ acknowledges the support of the one-hundred
talents program of  
the Chinese Academy of Sciences (CAS), the national science
foundation of China (grant No. 10821302 \& 10973027),  the CAS/SAFEA
International Partnership Program for  Creative Research Teams and the 973
program (grant No. 2007CB815401). AV is supported by the DOE at Fermilab. AV thanks the Fermilab Center for Particle Astrophysics for hospitality during the final stages of this work.


\bibliography{ShearBias}
\end{document}